\documentclass[prb,aps,floatfix]{revtex4}

\expandafter\let\csname equation*\endcsname\relax
\expandafter\let\csname endequation*\endcsname\relax
\usepackage{braket}
\usepackage{amsmath}

\usepackage[utf8]{inputenc}
\usepackage[dvips]{graphicx,graphics}
\usepackage{verbatim}
\newcommand{\D}{\mathrm{d}}
\renewcommand{\i}{\mathrm{i}}
\newcommand{\e}{\mathrm{e}}

\begin{document}

\title[Mean-field solution of the Hubbard model: the magnetic phase diagram]{Mean-field solution of the Hubbard model: the magnetic phase diagram}

\author{Y. Claveau, B. Arnaud, S. Di Matteo}

\affiliation{Groupe th\'eorie, D\'epartement Mat\'eriaux Nanosciences, Institut de Physique de Rennes UMR UR1-CNRS 6251, Universit\'e de Rennes 1, F-35042 Rennes Cedex, France}
\email{yann.claveau@univ-rennes1.fr, brice.arnaud@univ-rennes1.fr, sergio.dimatteo@univ-rennes1.fr}
\begin{abstract}
The present paper is based on our graduate lectures in condensed-matter physics. We found that the mean-field solution of the Hubbard model is an excellent tool to stimulate students' reflections towards the treatment of realistic magnetic interactions. We show by detailed analytical and numerical calculations how to find the mean-field solution of the model on a square lattice. We then interpret the physical implications of the ground-state magnetic phase diagram in terms of the electron density and the ratio between the Coulomb repulsion and the electron-structure bandwidth.
\end{abstract}

\maketitle

\section{Introduction}

In our graduate lectures in condensed-matter physics (second semester of master 1 or first semester of master 2) we found the mean-field solution of the Hubbard model a very useful tool for approaching realistic descriptions of materials. What is required is a general knowledge of the second-quantization formalism, with creation and annihilation operators that graduate students often find easier to visualize than the corresponding first-quantization wave functions. The mean-field solution of the Hubbard model is then obtained in a straigthforward way through a Fourier-transform to $\vec{k}$-space and a matrix diagonalisation. In spite of the relatively small amount of work, the lesson that a student can learn is very rich: he can construct by himself a magnetic phase diagram and understand in this way why ferromagnetism (FM) or antiferromagnetism (AFM) can be determined by the interplay of Coulomb repulsion, band energy and average electron density: an excellent way to start going beyond the independent-
electron approximation, towards the complexity of real materials.

Though the literature on the Hubbard model is vast, the model is usually dealt with only in the so-called two-pole approximation like in the original Hubbard papers \cite{hub1,hub2,hub3}, where the use of rather complex mathematical tools like Green-function equations-of-motion is mandatory. To the contrary, our mean-field solution allows dealing with continuity rather than discontinuity aspects compared to the usual single-particle approach: this might allow filling the gap between the latter and the more advanced research treatments of condensed-matter physics.

The present paper is organised as follows: in section 2 we introduce the Hubbard hamiltonian and our notation. Section 3 is devoted to the solution of the model in the mean-field approximation on a square lattice. We have chosen the square lattice in order to fix a realistic case (like, e.g., copper sites in CuO$_2$-planes of superconducting cuprates), by keeping at the same time a simple geometry. In section 4 we describe the computational details needed to obtain the ground-state phase diagram and discuss it with respect to the physical parameters of interest. Finally, in section 5 we linger on possible generalisations as a long-term exercise for students and draw our conclusions.

\section{The Hubbard model}
\subsection{Definitions}
The Hubbard Hamiltonian in the simplest case of non-degenerate band \cite{hub1}, i.e., with one orbital per site, can be expressed in the second-quantization formalism \cite{formalism} as: 

\begin{equation}\label{hubham}
{\hat{H}}_H=-\sum_{ij\sigma}t_{ij}{\hat{c}}^{\dagger}_{i\sigma}{\hat{c}}_{j\sigma} + U\sum_i {\hat{n}}_{i\uparrow}{\hat{n}}_{i\downarrow}\equiv {\hat{H}}_t + {\hat{H}}_U
\end{equation}

Here ${\hat{c}}^{\dagger}_{i\sigma}$ is as usual an operator representing the creation of an electron of spin $\sigma$ ($=\uparrow,\downarrow$) at site $i$, ${\hat{c}}_{j\sigma}$ the annihilation of an electron of spin $\sigma$ at site $j$ and $t_{ij}$ is the amplitude of the process, the so-called hopping amplitude from site $j$, where the electron is destroyed, to site $i$, where the electron is created. We finally notice that the sum over $i,j$ is unrestricted and $t_{ij}=(t_{ji})^*$, see Eq.~(\ref{hopping}): therefore the Hamiltonian is hermitian, as it should. The term $t_{ij}$ is the traduction in the second-quantization language of both the kinetic energy and the crystal-potential energy associated with an electron at site $i$:

\begin{equation}\label{hopping}
 t_{ij} \equiv t_{ij,\sigma} = \int \D^3\vec{r} \ \varphi^*_{\vec{R}_i,\sigma}(\vec{r}) \left( -\frac{\hbar^2\nabla^2}{2m} +V(\vec{r}) \right) \varphi_{\vec{R}_j,\sigma}(\vec{r})
\end{equation} 

Wannier wave-functions \cite{ashcroft} $\varphi_{\vec{R}_i,\sigma}(\vec{r})$ are centered at site $\vec{R}_i$. We suppose that the hopping amplitude does not depend on the spin variable and therefore we drop the $\sigma$ label. The term $V(\vec{r})$ represents the periodic crystal potential energy. As stated above, one approximation that is usually employed for the hopping term $t_{ij}$ is to consider it different from zero only when $i$ and $j$ are nearest-neighbour sites ($t_{ij}=t$), as their overlap is usually the largest. In this case the conventional minus sign associated to the hopping term in Eq.~(\ref{hubham}) allows having a minimum at the $\Gamma$-point in the reciprocal space ($\vec{k}=0$) for positive $t$.

Operators ${\hat{n}}_{i\sigma} \equiv {\hat{c}}^{\dagger}_{i\sigma}{\hat{c}}_{i\sigma}$ count the number of particles at site $i$ with spin $\sigma$. They are projection operators, ie, ${\hat{n}}^2_{i\sigma} = {\hat{n}}_{i\sigma}$ (either there is one electron with spin $\sigma$ at site $i$ or there are no electrons). The expectation value $\langle {\hat{n}}_{i\sigma}\rangle \equiv \langle\Psi_0|{\hat{n}}_{i\sigma}|\Psi_0\rangle$ in the many-body ground-state $|\Psi_0\rangle$ represents the electron density $n_{i\sigma}$ at site $i$ with spin $\sigma$ (mean occupation number). The physical origin of ${\hat{H}}_U$ is the Coulomb repulsion of the electrons: when at site $i$ both spin-up and spin-down electrons are present, from Eq.~(\ref{hubham}) they contribute to the total energy with a term $+U$, as both ${\hat{n}}_{i\uparrow}$ and ${\hat{n}}_{i\downarrow}$ in ${\hat{H}}_U$ give one. If, on the other side, the two electrons belong to two separate atoms, they do not feel any Coulomb repulsion (this is of 
course a strong constraint !). Formally, the on-site Coulomb repulsion can be written as:

\begin{eqnarray}\label{onsitecoulomb}
 U & \equiv \frac{e^2}{4\pi\varepsilon_0}\int \D^3\vec{r} \  \D^3\vec{r} \ ' \ \big|\varphi_{\vec{R}_i,\sigma}(\vec{r})\big|^2 \frac{1}{\big|\vec{r}-\vec{r}\,'\big|} \big|\varphi_{\vec{R}_i,\bar{\sigma}}(\vec{r} \ ')\big|^2 
\end{eqnarray} 

The Coulomb term does not depend on the site-label $i$, if we suppose the system homogeneous. 
Notice that in the extreme limit $U/t=0$, we recover a purely band-like (tight-binding) picture, with just kinetic energy and crystal periodic potential, and in the 
opposite limit $t/U=0$, we find a purely atomic picture.

\subsection{The square lattice}

\begin{figure}
 \begin{center}{
 \includegraphics[width=0.97\linewidth]{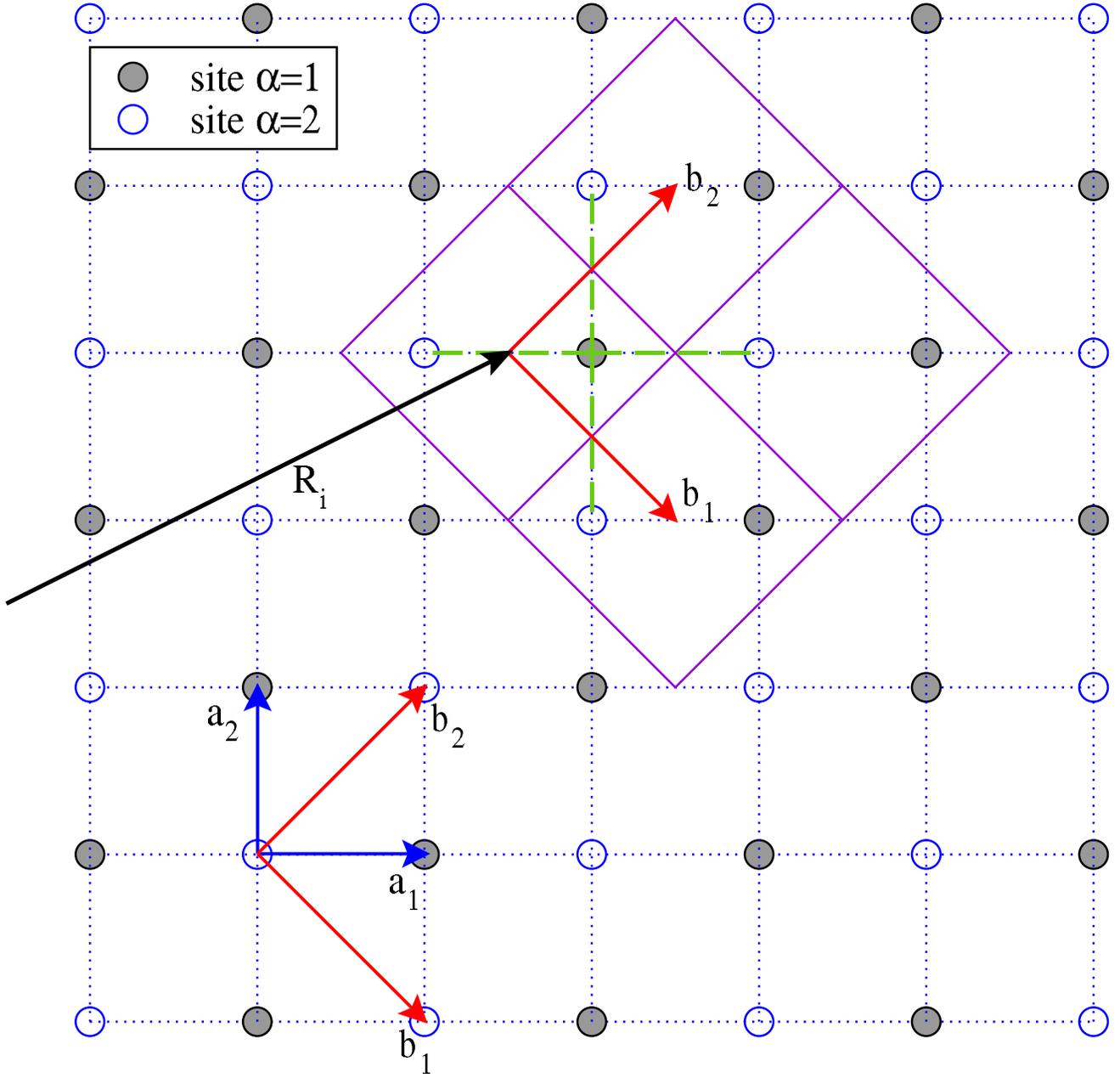}}
\caption{{Single ($\vec{a}_1$,$\vec{a}_2$) and double ($\vec{b}_1$,$\vec{b}_2$) cells for the square lattice. Nearest neighbours of $\alpha=1$, $\vec{R}_i$ are highlighted by the green dashed lines. The four violet cells represent nearest-neighbour cells in the ($\vec{b}_1$,$\vec{b}_2$)-basis.}}
\label{cell}
\end{center}
\end{figure}

The bi-dimensional square lattice is drawn in figure \ref{cell}. The unit-cell is spanned by the 2 vectors $\vec{a}_1$ and $\vec{a}_2$, of common length $a$. However, for our calculations we consider the cell of double area spanned by the 2 vectors $\vec{b}_1$ and $\vec{b}_2$, with the idea of looking for possible antiferromagnetic ground-states. Such a cell encloses two atomic sites that can in principle be inequivalent (e.g., spin $\uparrow$ and spin $\downarrow$). In what follows, we adopt the site-label $\vec{R}_i$ for the double unit cell and, within each cell, the two atoms are labeled by an extra index $\alpha = 1,2$. For example, as shown in figure \ref{cell}, nearest neighbours of $\alpha=1$ sites are necessarily $\alpha=2$ and vice-versa (bipartite lattice). The general creation (annihilation) operator for an electron at site $\vec{R}_i$, position $\alpha$, spin $\sigma$ can be written as: ${\hat{c}}^{\dagger}_{i\alpha\sigma}$ (${\hat{c}}_{i\alpha\sigma}$).
Equation (\ref{hubham}) should be changed accordingly:

\begin{equation}\label{hubhamalpha}
{\hat{H}}_{H}=-\sum_{ij\alpha\alpha'\sigma}t_{ij}^{\alpha\alpha'}{\hat{c}}^{\dagger}_{i\alpha\sigma}{\hat{c}}_{j\alpha'\sigma} + U\sum_{i\alpha} {\hat{n}}_{i\alpha\uparrow}{\hat{n}}_{i\alpha\downarrow}\equiv {\hat{H}}_{\tilde{t}} + {\hat{H}}_{\tilde{U}}
\end{equation}

\noindent The structure constants $t_{ij}^{\alpha\alpha'}$ for our problem (nearest-neighbour hopping) can be written as follows:

\begin{align}
 t_{ij}^{\alpha\alpha'} =&  - t\delta_{\alpha,1}\delta_{\alpha',2} \!\!\left[ \delta_{\vec{R}_i,\vec{R}_j} \!\!\!+ \delta_{\vec{R}_i+\vec{b}_1,\vec{R}_j} \!\!\!+ \delta_{\vec{R}_i+\vec{b}_2,\vec{R}_j} \!\!\!+ \delta_{\vec{R}_i+\vec{b}_1+\vec{b}_2,\vec{R}_j} \right] \nonumber \\
& - t\delta_{\alpha',1} \delta_{\alpha,2} \!\!\left[ \delta_{\vec{R}_i,\vec{R}_j}\!\!\! + \delta_{\vec{R}_i-\vec{b}_1,\vec{R}_j} \!\!\!+ \delta_{\vec{R}_i-\vec{b}_2,\vec{R}_j} \!\!\!+ \delta_{\vec{R}_i-\vec{b}_1-\vec{b}_2,\vec{R}_j} \right]
\label{struc_const}
\end{align}

In spite of the cumbersome form, the meaning of terms appearing in (\ref{struc_const}) is quite straightforward: the first line represents the four nearest-neighbour hopping energies (represented by dashed green lines in figure \ref{cell}) when site $\vec{R}_i$ is of $\alpha=1$-type and site $\vec{R}_j$ is of $\alpha=2$-type, and the second line the opposite case. As we employ the reciprocal lattice of the ($\vec{b}_1$,$\vec{b}_2$) direct lattice, we must express the nearest neighbours in terms of linear combinations of $\vec{b}_1$, $\vec{b}_2$ vectors. 

\section{Mean-field solution }

In order to derive the full mean-field solution, we first discuss the tight-binding solution, for $U=0$, and then analyse the action of the ${\hat{H}}_{\tilde{U}}$ term in mean-field. In both cases we can perform a Fourier transform to the $\vec{k}$-space, in order to gain full advantage of the 2-dimensional periodicity. 
As stated above, we use the reciprocal space of the ($\vec{b}_1$,$\vec{b}_2$) cell. 
The Fourier transform of the creation operator is: ${\hat{c}}^{\dagger}_{i\alpha\sigma} = \frac{1}{\sqrt{N}}\sum_{\vec{k}}\e^{-\i\vec{k}\cdot\vec{R}_i} {\hat{c}}^{\dagger}_{\vec{k}\alpha\sigma}$. That of the annihilation operator is the hermitian conjugate: ${\hat{c}}_{i\alpha\sigma} = \frac{1}{\sqrt{N}}\sum_{\vec{k}}\e^{\i\vec{k}\cdot\vec{R}_i} {\hat{c}}_{\vec{k}\alpha\sigma}$.
Here $N$ is the number of unit cells spanned by the vectors $\vec{b}_1$ and $\vec{b}_2$.

\subsection{Tight-binding solution: $U=0$}

Consider first the hopping part of the Hamiltonian, ${\hat{H}}_{\tilde{t}}$. By inserting the $\vec{k}$-transformed operators, we get the usual tight-binding band structure \cite{ashcroft}: 
 
\begin{align}\label{H0a}
{\hat{H}}_{\tilde{t}} & = -\frac{1}{N} \sum_{ij\alpha\alpha'\sigma} t_{ij}^{\alpha\alpha'}\sum_{\vec{k}\vec{k}'} \e^{-\i\vec{k}\cdot\vec{R}_i}{\hat{c}}^{\dagger}_{\vec{k}\alpha\sigma} \e^{\i\vec{k}'\cdot\vec{R}_j}{\hat{c}}_{\vec{k}'\alpha'\sigma} \nonumber \\
& = -\frac{1}{N}\sum_{\vec{k}\vec{k}'\alpha\alpha'\sigma}{\hat{c}}^{\dagger}_{\vec{k}\alpha\sigma}{\hat{c}}_{\vec{k}'\alpha'\sigma} \sum_{ij} t_{ij}^{\alpha\alpha'}\e^{-\i\vec{k}\cdot\vec{R}_i}\e^{\i\vec{k}'\cdot\vec{R}_j} \nonumber \\
& = -\sum_{\vec{k}\vec{k}'\alpha\alpha'\sigma}{\hat{c}}^{\dagger}_{\vec{k}\alpha\sigma}{\hat{c}}_{\vec{k}'\alpha'\sigma} \sum_{j}t_{\vec{\eta}_j}^{\alpha\alpha'}\e^{\i\vec{k}'\cdot\vec{\eta}_j} \frac{1}{N} \sum_i \e^{-\i(\vec{k}-\vec{k}')\cdot\vec{R}_i} \nonumber \\
& = \sum_{\vec{k}\alpha\alpha'\sigma}{\hat{c}}^{\dagger}_{\vec{k}\alpha\sigma}{\hat{c}}_{\vec{k}\alpha'\sigma} \varepsilon_{\vec{k}}^{\alpha\alpha'}
\end{align}

\noindent where we defined the matrix hamiltonian in the $\alpha$-$\alpha '$ basis as: $\varepsilon_{\vec{k}}^{\alpha\alpha'}= -\sum_{j}t_{\vec{\eta}_j}^{\alpha\alpha'}\e^{\i\vec{k}\cdot\vec{\eta}_j}$. In passing from the second to the third line of (\ref{H0a}) we used the translational invariance of the structure factor $t_{ij}^{\alpha\alpha'}$: this means that if we write $\vec{R}_j=\vec{R}_i+\vec{\eta}_{ij}$, the vectors $\vec{\eta}_{ij}$, and therefore the way of counting nearest neighbours, are independent of the starting point $\vec{R}_i$. So, we can write: $\vec{\eta}_{ij}\rightarrow \vec{\eta}_{j}$. Moreover we used the fact that $\frac{1}{N} \sum_i \e^{-\i(\vec{k}-\vec{k}')\cdot\vec{R}_i}=\delta_{\vec{k}\vec{k}'}$. 

\begin{figure}[!hbt]
 \begin{center}{\begin{minipage}{\linewidth}
 a)\includegraphics[width=0.97\linewidth]{bs.eps}
 b)\includegraphics[width=0.43\linewidth]{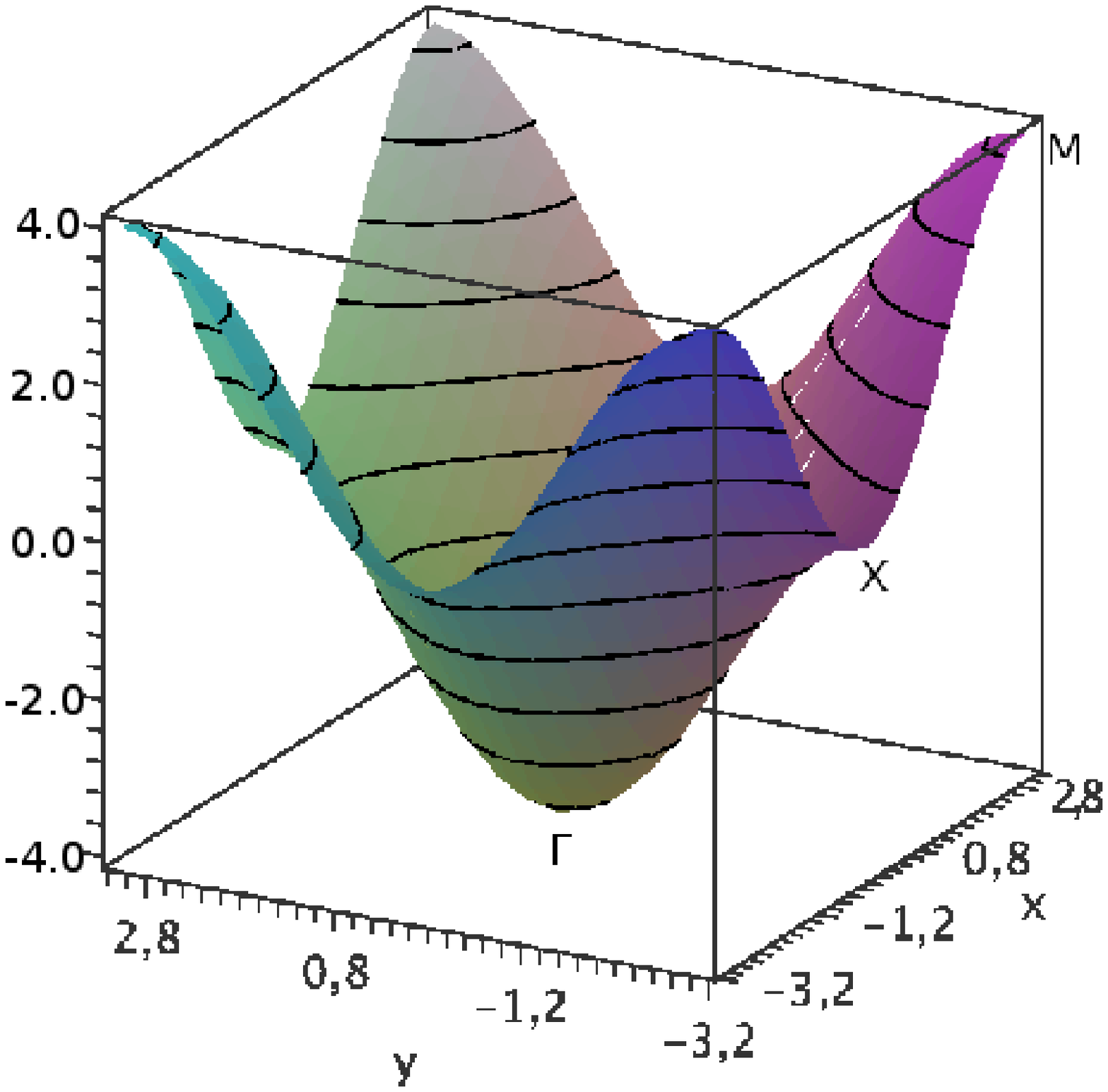} \includegraphics[width=0.43\linewidth]{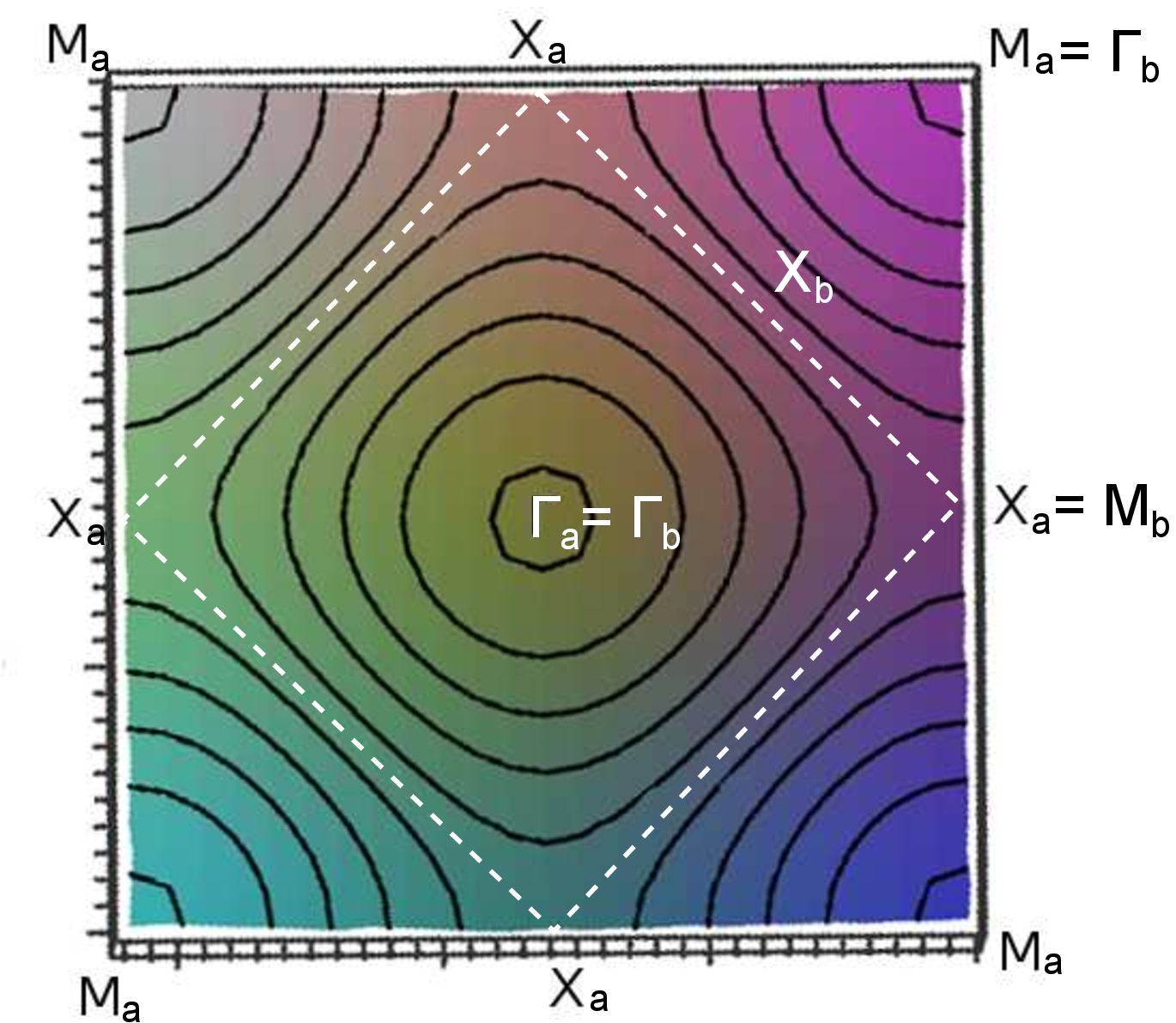}
 \end{minipage} }
\caption{{(a) Band structure in tight-binding approach ($U=0$) for single and double cells of figure \ref{cell}. For the double cell a second band appears due to the folding of the first band, as both points $\Gamma_a$ and $M_a$ of the reciprocal space of the single cell correspond to point $\Gamma_b$ of the reciprocal space of the double cell. (b) Full band-structure for the single cell. A Van-Hove singularity (the white dashed line which connects the $X_a$ points) appears at half filling.}}
\label{tb_band_structure}
\end{center}
\end{figure}

As we have two atoms per unit cell, the matrix hamiltonian is a 2x2 matrix ($\alpha,\alpha'=1,2$) that should be diagonalized in order to have the band structure $\varepsilon^{\pm}_{0\vec{k}}$.
From the explicit expression of the structure constants (\ref{struc_const}), we get the matrix: 
\begin{align}\label{H_02}
  H_{\tilde{t}}&=\left[ \begin{array}{cc}
              0    & t\gamma_{\vec{k}}  \\
              t\gamma^*_{\vec{k}} & 0
             \end{array}
      \right] 
\end{align}

\noindent where $\gamma_{\vec{k}}=- \left\{1+e^{-i\vec{k}\cdot(\vec{b}_1+\vec{b}_2)} +e^{-i\vec{k}\cdot\vec{b}_1}+e^{-i\vec{k}\cdot\vec{b}_2}\right\}$.
Its diagonalisation leads to two bands given by: 

\begin{align}\label{tbener}
\varepsilon^{\pm}_{0\vec{k}} = \pm t |\gamma_{\vec{k}}| = \pm t \left|2\cos\left(\vec{k}\cdot\frac{\vec{b}_1+\vec{b}_2}{2}\right)+2\cos\left(\vec{k}\cdot\frac{\vec{b}_2-\vec{b}_1}{2}\right)\right|
\end{align}

The band structure and density of states (DOS) per unit cell are depicted in figure \ref{tb_band_structure}, together with those for the single cell ($\vec{a}_1$,$\vec{a}_2$), for comparison. Bandwidth is $W=8t$. At half-filling, we have a nested Fermi surface, leading to a van Hove logarithmic singularity in the DOS \cite{ashcroft}. We remind that nesting is the property by which any point of the Fermi surface is related to another point of the Fermi surface by a fixed vector, in this case the vector $(\pi/a,\pi/a)$, because the Fermi surface is a square at half-filling, as shown in figure \ref{tb_band_structure}b. 
Clearly, at this level, changing the choice of the unit cell (a pure convention) does not change our results: in fact, the second band is just the folding of the first band of the single cell, as can be also seen by considering that $\frac{\vec{b}_1+\vec{b}_2}{2}=\vec{a}_1$ and $\frac{\vec{b}_2-\vec{b}_1}{2}=\vec{a}_2$. This comes from the fact that the point $M_a\equiv (\pi/a,\pi/a)$ in the reciprocal cell of the direct cell ($\vec{a}_1$,$\vec{a}_2$) becomes equivalent to $\Gamma_b\equiv (0,0)$ in the reciprocal cell of the direct cell ($\vec{b}_1$,$\vec{b}_2$). The two DOS of Fig.~\ref{tb_band_structure} are one the double of the other just because they are normalized per unit cell and there are two atoms per unit cell in the case of the double cell.

\subsection{Introduction of ${\hat{H}}_{\tilde{U}}$ in the tight-binding solution}

The presence of ${\hat{H}}_{\tilde{U}}$, equation (\ref{hubhamalpha}), changes these simple results even at a mean-field level.
We remind that the mean-field approximation corresponds to neglecting the fluctuations around the mean density. Such fluctuations are defined as $\Delta {\hat n}_{i\alpha\sigma} \equiv {\hat n}_{i\alpha\sigma} - \langle {\hat n}_{i\alpha\sigma}\rangle$, i.e., the difference between the exact number operator ${\hat n}_{i\alpha\sigma}$ and the mean occupation number $\langle {\hat n}_{i\alpha\sigma}\rangle$. From this relation, we get: ${\hat n}_{i\alpha\sigma} \equiv \langle {\hat n}_{i\alpha\sigma}\rangle + \Delta {\hat n}_{i\alpha\sigma}$. If we suppose homogeneity of the system, then $\langle {\hat n}_{i\alpha\sigma}\rangle$ is independent of the cell position $\vec{R}_i$, and we can drop the label $i$ and write:
\begin{align}
{\hat n}_{i\alpha\uparrow} {\hat n}_{i\alpha\downarrow} & = \left[ \Delta {\hat n}_{i\alpha\uparrow} + \langle {\hat n}_{\alpha\uparrow}\rangle\right] \cdot\left[ \Delta {\hat n}_{i\alpha\downarrow} + \langle {\hat n}_{\alpha\downarrow}\rangle \right] \\
 & = \Delta {\hat n}_{i\alpha\uparrow} \Delta {\hat n}_{i\alpha\downarrow} + \sum_{\sigma} {\hat n}_{i\alpha\sigma} \langle {\hat n}_{\alpha\bar{\sigma}} \rangle - \langle {\hat n}_{\alpha\uparrow}\rangle \langle {\hat n}_{\alpha\downarrow}\rangle   \nonumber
\end{align}
In a mean-field approximation we can neglect the first term of the previous equation, quadratic in the fluctuations. This implies that ${\hat{H}}_{\tilde{U}}$ becomes:
\begin{align}\label{meanf}
 {\hat{H}}^{\mathrm{MF}}_{\tilde{U}}\!=\! \underbrace{U \sum_{i\alpha\sigma} {\hat n}_{i\alpha\sigma} \langle n_{\alpha\bar{\sigma}} \rangle}_{\tilde{H}_U} - \underbrace{U N \sum_{\alpha} \langle n_{\alpha\uparrow}\rangle \langle n_{\alpha\downarrow}\rangle}_{E_U}
\end{align}
The second term $E_U$ is a constant for given magnetic configuration and number of particles, as it does not depend on creation or annihilation operators but only on their average values. However, it must be integrated in the calculation of the magnetic phase diagram, as such a term advantages paramagnetic configurations with respect to ferromagnetic and antiferromagnetic ($E_U$ is negative and the product $\langle n_{\alpha\uparrow}\rangle \langle n_{\alpha\downarrow}\rangle$, for fixed number of particles per site, is maximum when $\langle n_{\alpha\uparrow}\rangle = \langle n_{\alpha\downarrow}\rangle$). 

By using the Fourier transform as for equation (\ref{H0a}), the first term of (\ref{meanf}) becomes: $\tilde{H}_U = U \sum_{i\alpha\sigma} \langle n_{\alpha\bar{\sigma}}\rangle {\hat{c}}^{\dagger}_{i\alpha\sigma} {\hat{c}}_{i\alpha\sigma}=U \sum_{\vec{k}\alpha\sigma} \langle n_{\alpha\bar{\sigma}}\rangle {\hat{c}}^{\dagger}_{\vec{k}\alpha\sigma} {\hat{c}}_{\vec{k}\alpha\sigma}$, diagonal in $\alpha$.
Therefore the energy per $\vec{k}$-point for a spin-$\sigma$ electron is obtained by diagonalizing the 2x2 matrix:
\begin{align}\label{h_tot}
&\left[ \begin{array}{cc}
U\langle n_{1\bar{\sigma}}\rangle &  t\gamma_{\vec{k}}  \\
t\gamma^*_{\vec{k}}                                & U\langle n_{2\bar{\sigma}} \rangle
        \end{array}
      \right] 
\end{align}
Where $\gamma_{\vec{k}}$ has been defined after (\ref{H_02}). 
This Hamiltonian can be easily diagonalized analytically, as the associated eigenvalue problem leads to the second-order algebraic equation for the eigenenergies $\varepsilon^{\pm}_{\vec{k}}$:
\begin{align}\label{eq_ham}
 \varepsilon^{2}_{\vec{k}} - \varepsilon_{\vec{k}} \left(U\langle n_{1\bar{\sigma}}\rangle+U\langle n_{2\bar{\sigma}}\rangle\right) + U^2 \langle n_{1\bar{\sigma}}\rangle \langle n_{2\bar{\sigma}}\rangle-t^2|\gamma_{\vec{k}}|^2 =0 
\end{align}

The solutions of this second-order equation provide the mean-field band-energies per $\vec{k}$-point for a spin-$\sigma$ electron, once we add again the term $E_U$ (divided by $N$, because $\varepsilon_{\vec{k}}$ is the energy per spin-$\sigma$ electron):

\begin{align}\label{sol_ham}
& \varepsilon^{\pm}_{\vec{k}\sigma} = U\left(\frac{\langle n_{1\bar{\sigma}}\rangle+\langle n_{2\bar{\sigma}}\rangle}{2}\right) -U \big(\langle n_{1\uparrow}\rangle \langle n_{1\downarrow}\rangle + \langle n_{2\uparrow}\rangle \langle n_{2\downarrow}\rangle \big)  \\
& \pm \frac{1}{2}\textstyle{\sqrt{U^2(\langle n_{1\bar{\sigma}}\rangle-\langle n_{2\bar{\sigma}}\rangle)^2+16t^2\left(\cos\left[\vec{k}\cdot\frac{\vec{b}_1+\vec{b}_2}{2}\right]+\cos\left[\vec{k}\!\cdot\frac{\vec{b}_1-\vec{b}_2}{2}\right]\right)^2}}  \nonumber \\
& = U\left(\frac{\langle n_{1\bar{\sigma}}\rangle+\langle n_{2\bar{\sigma}}\rangle}{2}\right) -U \big(\langle n_{1\uparrow}\rangle \langle n_{1\downarrow}\rangle + \langle n_{2\uparrow}\rangle \langle n_{2\downarrow}\rangle \big) \nonumber \\
& \pm \frac{1}{2}\textstyle{\sqrt{U^2(\langle n_{1\bar{\sigma}}\rangle-\langle n_{2\bar{\sigma}}\rangle)^2+4 (\varepsilon^{\pm}_{0\vec{k}})^2}}   \nonumber
\end{align}

From the last line we see that ${\hat{H}}^{MF}_{\tilde{U}}$ contributes to the total energy as a density-dependent and spin-dependent (because of $\bar{\sigma}$) renormalization of the tight-binding energy $\varepsilon^{\pm}_{0\vec{k}}$. This double dependence on the density and on the spin is at the basis of the richness in the phase diagram shown in figure \ref{phase_diagram}, allowing to get paramagnetic (PM), ferromagnetic (FM) and antiferromagnetic (AFM) phases.

\section{Results and discussion}

\subsection{Computational details.}

\begin{figure}[!hbt]
 \begin{center}{
 \includegraphics[width=0.9\linewidth]{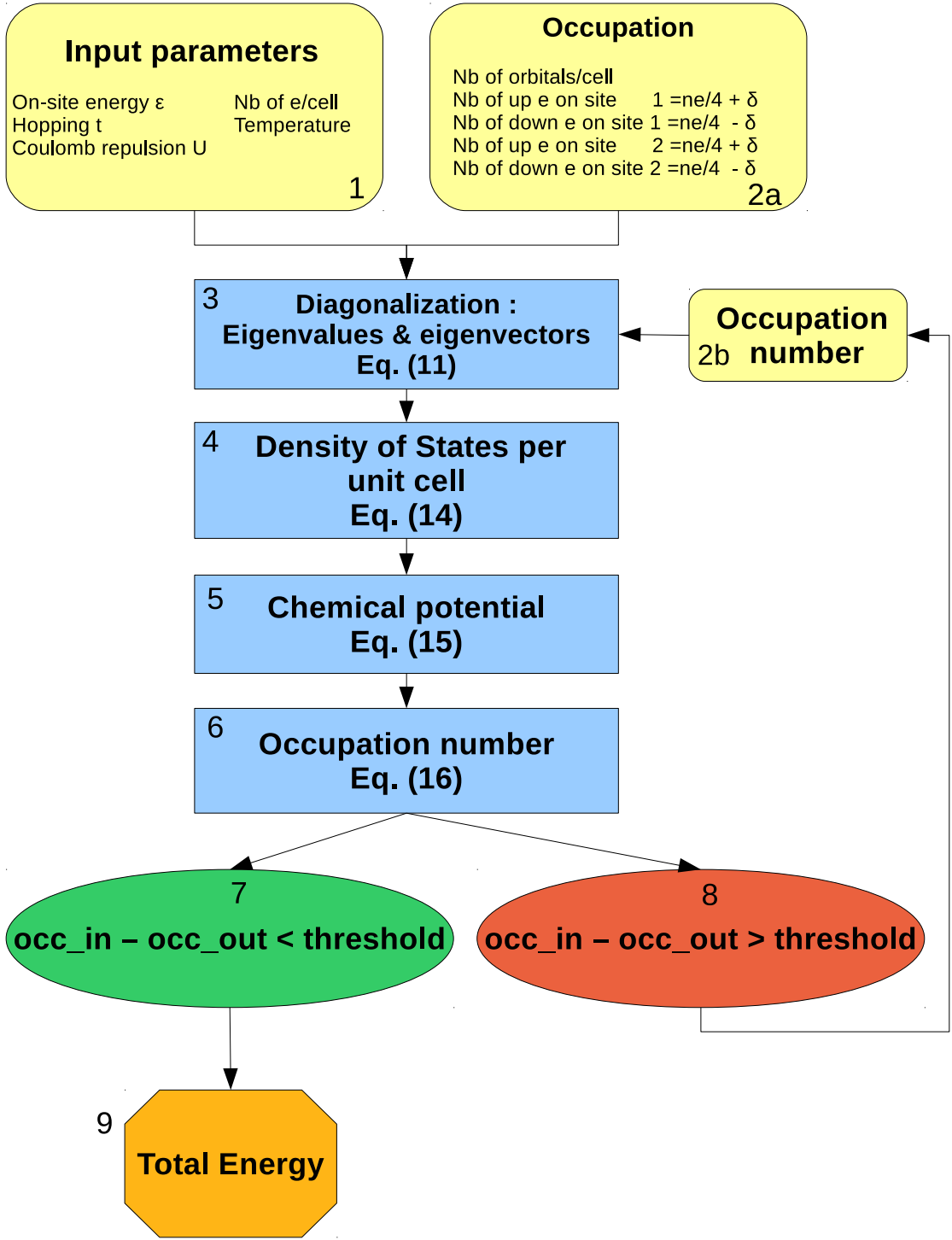}}
\caption{{Scheme of the self-consistent algorithm leading to the ground-state phase diagram of figure \ref{phase_diagram}. Equations are described in the text}}
\label{algo}
\end{center}
\end{figure}

In order to find the ground-state energy for a given magnetic configuration, we need to employ a self-consistent scheme as, for a given spin $\sigma$, the energy depends on the mean occupation number of opposite spin, $\langle {\hat n}_{\alpha\bar{\sigma}}\rangle$, which in turns depends on the eigenvectors of the energy matrix itself.
As shown in figure \ref{algo}, we therefore start from a given configuration of input parameters (frames 1 and 2), we follow steps 3, 4, 5 and 6, up to the self-consistency condition expressed in frames 7 and 8, and finally move back to frame 2 or end to frame 9, depending on whether the condition is not satisfied or satisfied, respectively. The condition is satisfied when the output occupation number is the same as the input occupation number within a threshold that we fixed to $10^{-5}$.   

In more details, we proceeded as follows: for practical reasons we found it simpler to numerically diagonalize the Hamiltonian of (\ref{h_tot}) through the Lapack libraries \cite{lapack} (frame 3), in order to determine  eigenvalues $\varepsilon_{j\sigma}(\vec{k})$ and eigenvectors $\Ket{\Psi^{\,j}_{\vec{k}\sigma}}=\sum_{\alpha}A^{\,j}_{\alpha,\sigma}(\vec{k}) \Ket{\alpha\vec{k}}$ (with $j=1,2$). We actually work with the chemical potential in order to fix the particle density aposteriori: ${\hat H}_H^{\mu}={\hat H}_H-\mu \sum_{i\alpha\sigma}{\hat n}_{i\alpha\sigma}$. 
The calculations performed in frames 4, 5 and 6 are based on equations (\ref{DOS}), (\ref{chemical_potential}) and (\ref{occupation}) given below. The first of these equations expresses the DOS per unit cell, $\rho(\varepsilon)$:
\begin{equation}\label{DOS}
 \rho(\varepsilon)=\frac{1}{N}\sum_{{\vec{k}}, j, \sigma} 
\delta\left[\varepsilon-\varepsilon_{j \sigma}({\vec{k}})\right]
\end{equation}

The Dirac delta function in (\ref{DOS}) is numerically calculated through a Gaussian function whose broadening is optimized after a convergence study: if the width of the Gaussian is too wide compared to the step between two energy points in $\varepsilon$, the DOS is too smooth, whereas if the width is too narrow, artificial oscillations appear. A more elegant approach might be to use the Methfessel-Paxton method \cite{Methfessel_Paxton}, usually implemented in more advanced packages for electronic structure calculations.

As a second step we determine the chemical potential $\mu$ by the implicit equation: 
\begin{equation}\label{chemical_potential}
\int_{-\infty}^{+\infty} \D\varepsilon \,\rho(\varepsilon) 
\frac{1}{\exp[\beta(\varepsilon-\mu)]+1}=n_e
\end{equation} 
Here $\beta=1/k_BT$ is the Boltzmann factor and we have defined the total number of electrons per unit cell (as used in figures \ref{algo}, \ref{hubbard_bs}, \ref{phase_diagram}): $n_e \equiv \sum_{\alpha, \sigma}\langle n_{\alpha\sigma} \rangle$. Finally, the average number of particles per site $\alpha$ and per spin $\sigma$ at a given temperature $T$ is given by:
\begin{equation}\label{occupation}
\langle n_{\alpha\sigma} \rangle=\frac{1}{N} \sum_{{\vec{k}}, j=1,2}
|A_{\alpha, \sigma}^{\,j}({\vec{k}})|^2
\frac{1}{\exp[\beta(\varepsilon_{j\sigma}({\vec{k}})-\mu)]+1}
\end{equation}

At finite temperature, the thermodynamic variable to minimize is of course not the total energy, $E$, but the free energy, $F=E-TS$, what implies a calculation of the entropy of the system, through the equation:
\begin{equation}
  S(T) = - k_B \int_{-\infty}^{\infty} \D\varepsilon \rho(\varepsilon) \Big\{ f( \varepsilon ) \ln[f(\varepsilon)] \, + \, [1-f(\varepsilon)] \ln[1-f(\varepsilon)] \Big\} 
\end{equation}

Here $f( \varepsilon )$ is the Fermi-Dirac distribution. 
However, in what follows, we are interested in the ground-state phase diagram of the model, i.e., at $T=0$. We minimized therefore the total energy and used the parameter $T$ of equations (\ref{chemical_potential}) and (\ref{occupation}) for convergence purposes only. It is in fact common practice to use a Fermi-Dirac distribution, characterized by a parameter $T$ different from zero, even in the $T\rightarrow 0$ limit, in order to smooth the step function that appears in equations (\ref{chemical_potential}) and (\ref{occupation}) for $T=0$.

\begin{figure}[!hbt]
 \begin{center}{
 \includegraphics[width=0.97\linewidth]{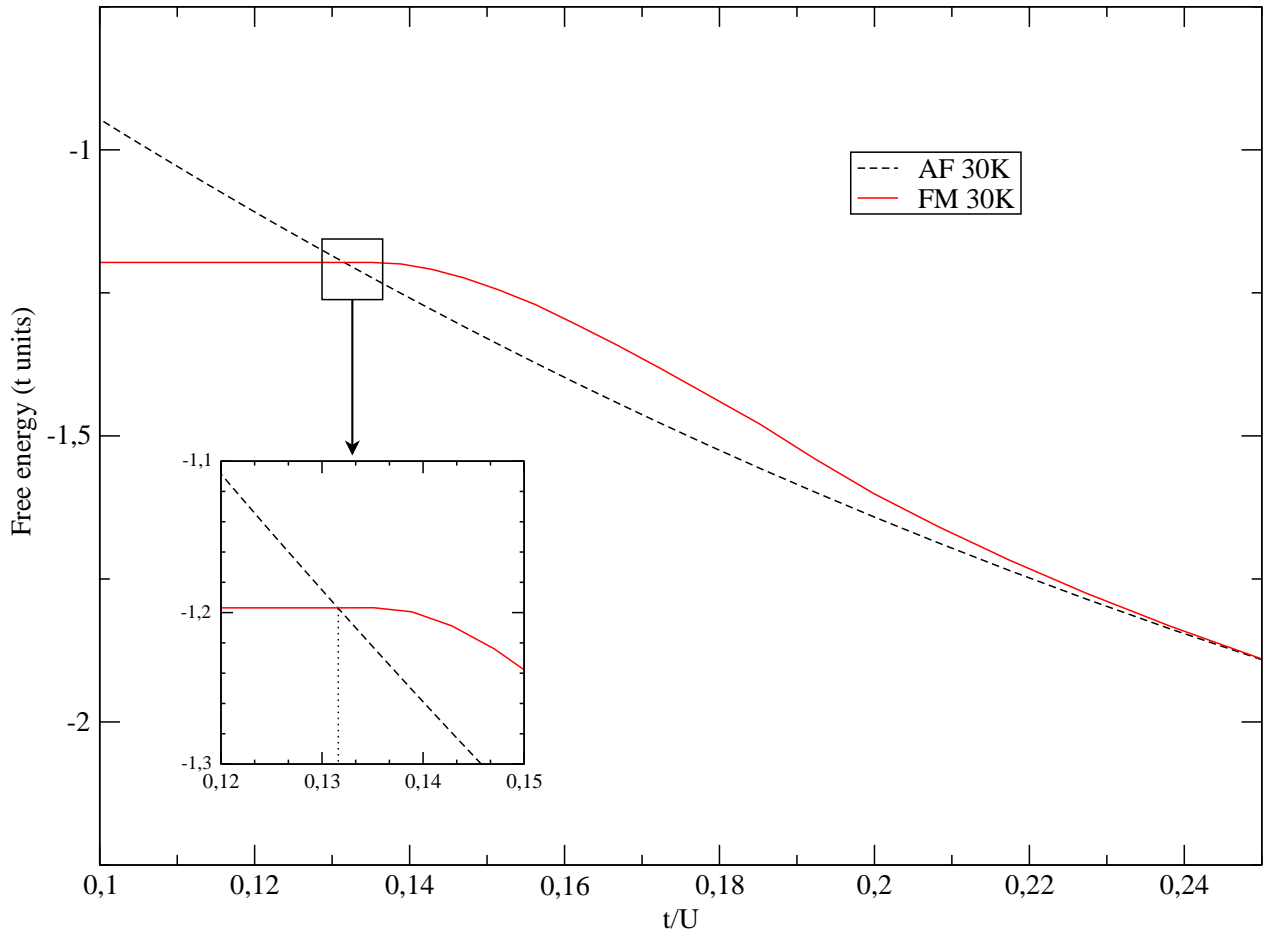}}
\caption{{A phase transition occurs where magnetic-energy curves cross each other. For $n$=0.8 the system becomes ferromagnetic below $t/U=0.13$. We have chosen an inverse temperature $\beta\sim 0.003$ eV.}}
\label{phase_transition}
\end{center}
\end{figure}

In general, we computed the total energy for each magnetic phase as a function of $t/U$. A phase transition is then characterized by the crossing of two (or more) free-energy curves: for example, we have represented the magnetic free energy as a function of $t/U$ in figure \ref{phase_transition} for a specific occupation number ($n = 0.8$), in order to highlight a magnetic (AFM/FM) transition. In this case the phase transition from AFM to FM phase is obtained at $t/U=0.13$, value at which the total FM energy becomes smaller than the AFM energy.
  
\subsection{Discussion of results: phase diagram.}

\begin{figure}[!hbt]
 \begin{center}
 \includegraphics[width=\linewidth]{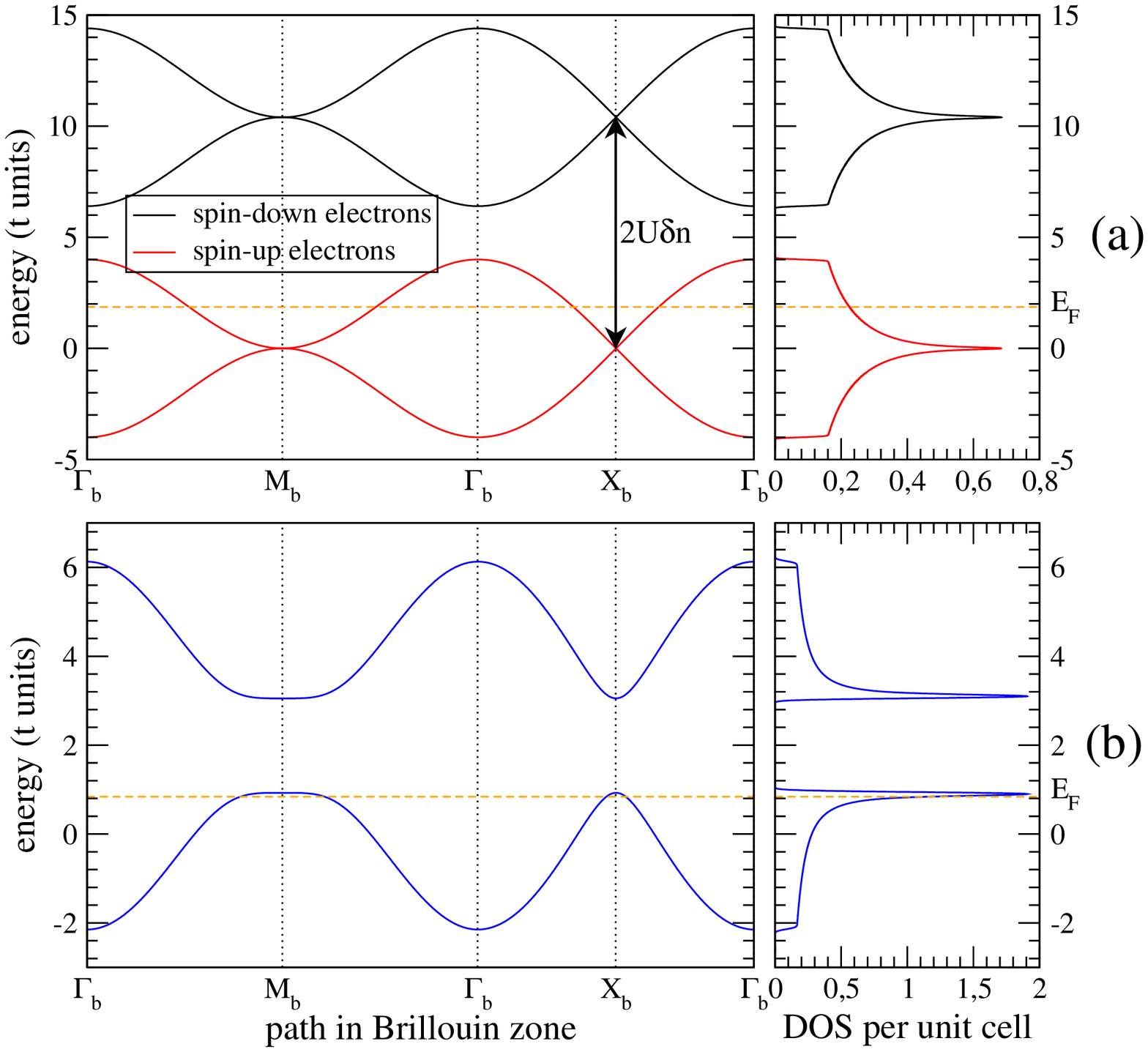}
 \caption{Band structure and DOS for: (a) FM configuration: $t/U=0.077$ ; $n_e=1.6$ ($n_{1\uparrow}=n_{2\uparrow}=0.8$ and $n_{1\downarrow}=n_{2\downarrow}=0$) and (b) AFM configuration: $t/U=0.2$ ; $n_e=1.6$ ($n_{1\uparrow}=n_{2\downarrow}=0.62$ and $n_{1\downarrow}=n_{2\uparrow}=0.18$). In the FM case the exchange splitting is $2U\delta n$. In the AFM case a Slater gap appears, leading to an insulator for $n_e=2.0$. The DOS has been obtained by modeling equation (\ref{DOS}) with a normalized Gaussian of width 0.05 $t$ and with a $500\times500$ $k$-point grid. }
\label{hubbard_bs}
\end{center}
\end{figure}

In figure \ref{hubbard_bs}, we have drawn the band structures for both FM and AFM phases to highlight the differences with the PM case. FM bands are shifted rigidly with respect to PM bands by about $\pm U\delta n$, where $\delta n$ is half the spin unbalance (see below). AFM bands are instead characterized by the opening of a gap that can be related to the difference $U(| \langle n_{1\bar{\sigma}}\rangle\!-\!\langle n_{2\bar{\sigma}}\rangle |)$ in (\ref{sol_ham}). This gap leads to an insulating phase for $n_e=2.0$. It is important to notice that such a gap is not related to a metal-insulator transition (MIT) of the Mott-Hubbard kind, as it is not related to the electronic correlations (that are absent by definition in a mean-field calculation), but to magnetism. This MIT is rather called Slater MIT, in honor of J.C. Slater that foresaw it in 1951 \cite{slater}. In fact, differently from Mott that did not originally ascribed his MIT to magnetic interactions, Slater thought that the origin of the metal-to-
insulator transition was determined by the onset of 
AFM long-range order, exactly as in the scheme described in the present paper. Therefore a Slater insulator is characterized by a band gap determined by a superlattice modulation of the magnetic periodicity. This is not the case of a Mott insulator.  

\vspace{0.5cm}

\begin{figure}[!hbt]
 \begin{center}{
 \includegraphics[width=0.97\linewidth]{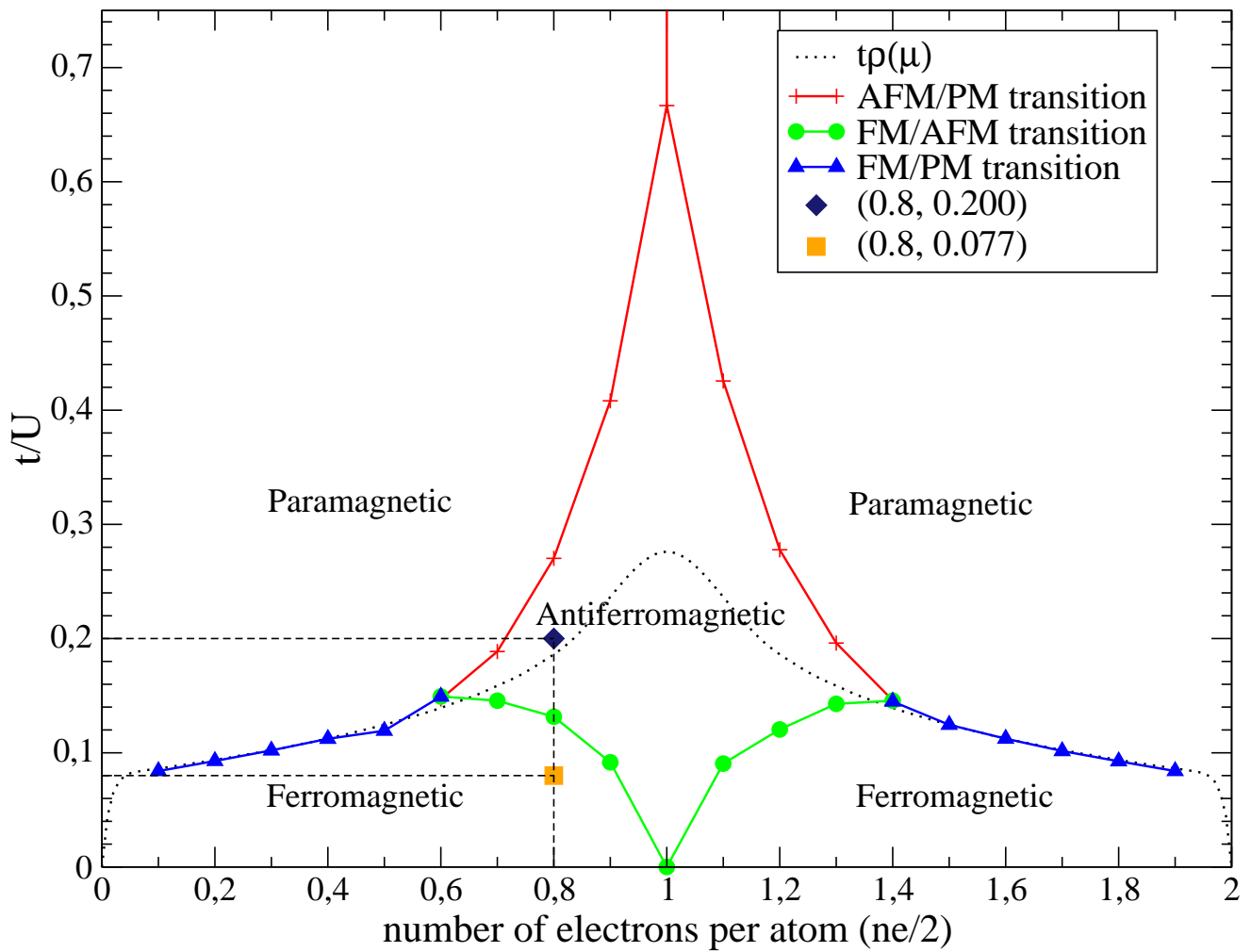}}
\caption{{Ground-state phase diagram of the Hubbard model on a square lattice as a function of the ratio $t/U$ and of the electron filling. The physical relevance of the curve $t\rho(\mu)$ is detailed in the text. We used an inverse-temperature value $\beta = 0.03 t$ in the equations. The two points (0.8, 0.077) and (0.8, 0.2) correspond to the band structure depicted in figure \ref{hubbard_bs}.}}
\label{phase_diagram}
\end{center}
\end{figure}

Our main result is the ground-state phase diagram, drawn in Figure \ref{phase_diagram} as a function of the number of electrons per site and of $t/U$. Such a phase diagram had been already obtained in the literature in 1985 by Hirsch \cite{hirsch}, though in a different context and without providing all the details of the derivation that can be found here.
As a general feature the phase diagram shows a clear symmetry around half filling, i.e., one electron per site, where antiferromagnetism is the lowest-energy configuration. This symmetry had to be expected, since it is a symmetry of the Hubbard hamiltonian for nearest-neighbour hopping. Far from half-filling, paramagnetism is advantaged by a high value of $t/U$, whereas a low value of $t/U$ leads instead to ferromagnetism. This tendency for the PM/FM phases can be easily understood: the ground-state of $n$ non-interacting electrons ($U=0$) is PM because the minimum-energy constraint in combination with the Pauli principle forces to fill all the energy levels from the lowest ($\varepsilon_{\rm{min}}$) to the highest ($\mu$) with an equal number of $n/2$ up and $n/2$ down electrons. In the opposite extreme case, for $U/t \rightarrow \infty$, the system can gain energy by a total magnetization (say, all electrons with spin up), in order to minimize the energy term $U \langle n_{\downarrow}\rangle$. In the 
intermediate $t/U$ cases, only the numerical study of equations (\ref{h_tot}), (\ref{DOS}), (\ref{chemical_potential}) and (\ref{occupation}) can provide us with the magnetic phase of the system. 

\begin{figure}[!hbt]
 \begin{center}{
 \includegraphics[width=0.97\linewidth]{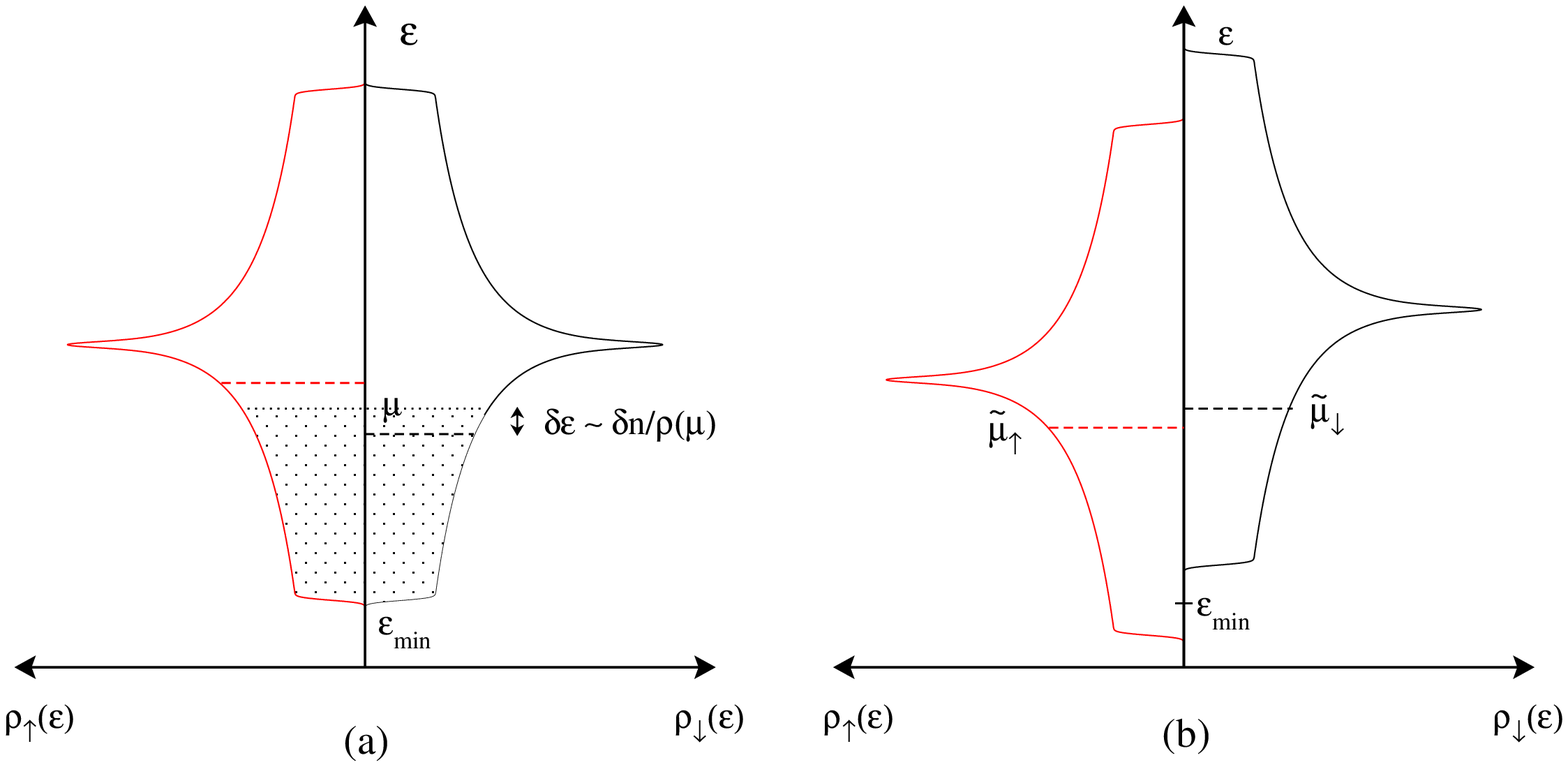}}
\caption{(a) If $\delta n \ll n$ spin-down electrons are moved to spin-up DOS, for $U=0$, this leads to a shift of the chemical potential $\pm\delta \varepsilon \sim \pm\delta n/\rho(\mu)$: PM state is always favored.
(b) If $U \neq 0$, then spin-up DOS is lowered in enegy by $-U\delta n + U(\delta n)^2/n$ and spin down DOS is raised in energy by $U\delta n + U(\delta n)^2/n$. Therefore, depending on $\rho(\mu)$, FM stability can be obtained for sufficiently high $U$ (see text).}
\label{stoner}
\end{center}
\end{figure}

There exists however a criterion that allows us to foresee the stability of the PM phase versus the FM phase just on the base of the two parameters $U$ and $\rho(\mu)$, the DOS at the Fermi energy ($\varepsilon_F = \mu$ at $T=0$). This is Stoner criterion \cite{Stoner}.
Start from the PM phase, represented in grey in  figure \ref{stoner}(a) ($n_{\uparrow}=n_{\downarrow}=n/2$) and move $\delta n$ electrons from spin-down to spin-up states, so that $n_{\uparrow}=n/2+\delta n$ and $n_{\downarrow}=n/2-\delta n$. This leads to a shift of the chemical potential by $\pm\delta \varepsilon \simeq \pm\delta n/\rho(\mu)$ (if $\delta n \ll n$). The system is not at equilibrium, as $\mu_{\uparrow}=\mu + \delta n/\rho(\mu)\neq \mu_{\downarrow}= \mu - \delta n/\rho(\mu)$. Yet, because of the change in $\delta n$ and of the mean-field Coulomb energy (\ref{meanf}), the minority-spin band shifts as a whole upward by $U\delta n + U (\delta n)^2/n$ and majority-spin band shifts as a whole downwards by $U\delta n - U (\delta n)^2/n$ as shown in figure \ref{stoner}(b). 

In this situation, if  ${\tilde {\mu}_{\uparrow}} = \mu + \delta n/\rho(\mu)-U\delta n + U (\delta n)^2/n$ is less than ${\tilde {\mu}_{\downarrow}}=\mu - \delta n/\rho(\mu)+U\delta n + U (\delta n)^2/n$, then it is favorable to still increase the number of spin-up electrons until ${\tilde {\mu}_{\uparrow}} = {\tilde {\mu}_{\downarrow}}$. Therefore, ferromagnetism appears when ${\tilde {\mu}_{\uparrow}} \leq {\tilde {\mu}_{\downarrow}}$. This leads to the Stoner criterion for ferromagnetic stability: 
\begin{equation} \label{stonereq}
U\rho(\mu) \geq 1
\end{equation}

Equation \ref{stonereq} can also be written as $\frac{t}{U} \leq t \rho(\mu)$. The corresponding equality, that marks the phase transition, has been reproduced in figure \ref{phase_diagram}. All our calculated points for the PM/FM transition lie on the theoretical curve $\frac{t}{U} = t \rho(\mu)$ represented by a dotted line in Figure \ref{phase_diagram}. We infer from the Stoner criterion that a ferromagnetic instability is expected in materials showing a high density of states at Fermi level. This is indeed the case for Fe, Co and Ni \cite{blundell}.
It is also possible to find a similar criterion for the AFM/PM and AFM/FM transitions, but its derivation is technically more involved because it is based on a Bogoliubov transformation. This leads to a gap-equation formally equivalent to the one of the BCS theory of superconductivity. A rather detailed description of this generalization can be found in \cite{Mahan}.

\section{Generalizations and conclusion.}

In the previous section we have analyzed formalism and phase diagram of the one-band Hubbard model on a square lattice in the mean-field approximation. For completeness, in this section we give a brief overview of four possible generalizations of the model to provide the link to the more advanced literature, with applications to real materials.

The first modification that can be dealt with concerns the application of (\ref{hubham}) to a different lattice. This primarily leads to a different DOS than the one shown in figure \ref{tb_band_structure}, thereby modifying quantitatively, but not necessarily qualitatively, the phase diagram. The fact that quantitative changes in the DOS do not necessarily imply qualitative (topological) modifications of the phase diagram can be understood by comparing our case with the example given in \cite{Mahan}, where a free-electron-gas parabolic DOS is employed: the resulting phase diagram is topologically similar to ours. Qualitatively similar behavior is also obtained in all cases where bipartite lattices are considered, as the honeycomb lattice. We remind that a lattice is called bipartite if two atoms of kinds A and B can be accommodated in it in such a way that any atom of kind A is only surrounded by atoms of kind B and vice-versa. On the contrary, qualitatively and quantitatively 
different results are obtained in the case of frustrated lattices, like the triangular lattice. This is the case because antiferromagnetic interactions can be depleted by geometrical frustrations and paramagnetism is generally advantaged \cite{lee} in lattices that are not bipartite.

As a second modification, it is possible to extend the hopping term beyond nearest neighbours. For example next-nearest-neighbour hopping is not necessarily zero as supposed in the present paper. In the square lattice such a term corresponds to an hopping integral between the two atoms along the directions of $\vec{b}_1$ and $\vec{b}_2$ of figure \ref{cell}. The effect of this term in the DOS of a square lattice is to remove the nesting property of Fermi surface at half filling. For a 2-dimensional square lattice such a calculation in the mean-field approximation has been performed, e.g., by Hirsch \cite{hirsch2}, who indeed found a deformed phase diagram with respect to that of figure \ref{phase_diagram}, without the symmetry around half-filling. 

In third place, instead of 2-dimensional systems we could move to 3-dimensional lattices. In the case of a cubic lattice, for example, this leads to the removal of the van Hove singularity, determined by the 2-dimensional square-lattice topology with nearest neighbours, and therefore to the removal of the logarithmic singularity at half-filling in the DOS of figure \ref{tb_band_structure}. All these modifications can of course be combined together, to get a final phase diagram that can be substantially different from the one presented in this paper even in the mean-field approximation. 

One final modification that applies to realistic systems, is to introduce multi-orbital Hubbard models and/or multi-band Hubbard models. Models where $d$ or $f$ orbitals are introduced belong to the first kind. In this case a further index $m$ up to 5 for $d$ orbitals and up to 7 for $f$ orbitals must be introduced to deal with electron creation and annihilation operators for different wave-functions (e.g., $d_{xy}$, $d_{yz}$, $d_{x^2-y^2}$, etc.). Hopping terms are then modified in a similar way as when we moved from (1) to (4). However, the extra labels, $\alpha$, $\alpha'$ in (4) and $m$, $m'$ in (\ref{hubhamorb}) below have different physical interpretations, $m$, $m'$ representing two different $d$ or $f$ (or sometimes $p$) orbitals on the same atom. The multiorbital Hubbard hamiltonian is written (see, e.g., \cite{hub2} or section II of \cite{prbv2o3}):

\begin{align}\label{hubhamorb}
& H=\sum_{ijmm'\sigma}t_{ij}^{mm'}{\hat{c}}^{\dagger}_{im\sigma}{\hat{c}}_{jm'\sigma} + \sum_{imm'\sigma\sigma'} U^{mm'} {\hat{n}}_{im\sigma}{\hat{n}}_{im'\sigma'} \nonumber \\
& +J\sum_{m \neq m'} \big( {\hat{c}}^{\dagger}_{im\uparrow}{\hat{c}}^{\dagger}_{im\downarrow} {\hat{c}}_{im'\downarrow}{\hat{c}}_{im'\uparrow} - {\hat{c}}^{\dagger}_{im\uparrow}{\hat{c}}_{im\downarrow} {\hat{c}}^{\dagger}_{im'\downarrow}{\hat{c}}_{im'\uparrow}\big)
\end{align}

It is important to underline that in this case, several intra-atomic Coulomb terms appear, depending on whether intra-orbital ($U^{mm'}$, with $m=m'$) or inter-orbital ($U^{mm'}$, with $m\neq m'$) Coulomb repulsion is concerned. Moreover, because of the multi-dimensional orbital degree of freedom, also Hund's exchange $J$ appears, for $m \neq m'$. Interestingly, this implies the appearence of an exchange term in the mean-field approximation: the Hartree approximation of this paper would become an Hartree-Fock approximation. 

Finally, multi-band Hubbard models are those where several atomic species are present, not all necessarily characterized by the same Hubbard $U$ (that can also be zero in some cases). Probably the most famous of this kind are the Anderson periodic model \cite{anderson}, used to describe the interaction of a localised electron (e.g. an $f$ electron) with a 'Fermi sea', or the $pd$-model \cite{emery,varma} used to describe CuO$_2$-planes in superconducting cuprates, where 2 kinds of $p$ bands and 1 $d$ band are introduced.

This rapid oveview shows the potential applications that the generalisation of a simple mean-field solution of the Hubbard model can have. We would like to stress, again, that many of these generalisations \cite{hirsch,lee,hirsch2,prbv2o3,anderson,emery,varma} are not just academic exercises and can bring the interested student very close to real researches in condensed-matter physics. At the same time, based on our experience, we found out that the calculations and the physical concepts presented in this paper are in average understood by graduate students. Last but not least, the self-consistent numerical procedure used in section 4 to diagonalise the hamiltonian and find the phase diagram can represent a useful tool for students to make the link between formal implicit formulas and the way useful figures have to be derived. 

As a final remark, we should also remind that the recent literature about the Hubbard model is extremely vast and, in the last thirty years, many developments have been undertaken in strong sinergy with the discovery of new strongly-correlated electron materials, mainly, transition-metal oxides \cite{olesxxx}: an excellent review on this aspect is \cite{imada}. A book where the analytical properties of the Hubbard model are developed, especially in relation with the so-called two-pole approximation (or expansion around the atomic limit) is \cite{ovchin}: the approach is rather mathematical and many of the exact results of the model, together with the original references, are discussed in detail. Concerning numerical approaches, related to band-structure calculations, nowadays no PhD (and post-doc) student should neglect the study of the two methods (in order of difficulty) known as DFT+U (Density Functional Theory + Hubbard U) \cite{anisimov1,anisimov2,cococcioni} and DFT+DMFT (Density Functional Theory + 
Dynamical Mean-Field Theory) \cite{kotliar,kotliar2}. DFT+U is based on the mean-field approach developed in the present paper, with the only difference of being orbitally and site-dependent. Dynamical mean-field approaches, instead, allow a more complex dynamics of the system, with shifts of the spectral weight and non-Fermi-liquid behaviour: we refer to the comprehensive review work \cite{kotliar} for a deeper introduction.

\vspace{0.5cm}

\bibliography{Hubbard_EuropJ}

%\subsection{Acknowledgments}
%Authors wishing to acknowledge assistance or encouragement from 

%\section*{References}
%\begin{thebibliography}{10}
%\bibitem{book1} Goosens M, Rahtz S and Mittelbach F 1997 {\it The \LaTeX\ Graphics Companion\/} 
%(Reading, MA: Addison-Wesley)
%\bibitem{eps} Reckdahl K 1997 {\it Using Imported Graphics in \LaTeX\ } (search CTAN for the file `epslatex.pdf')
%\end{thebibliography}

\end{document}